%% file: Free-new.tex
\input phyzzx\input mydef
\overfullrule=0pt
\date{November, 2020}
\date{November, 2020}
\titlepage
\title{On the Free Energy of Solvable lattice Models}
\author{Doron Gepner}
\vskip20pt
\line{\it\hfill  Department of Particle Physics and Astrophysics, Weizmann Institute,\hfill}
 \line{\it\hfill Rehovot 76100,  Israel\hfill} 
 
 \abstract
 
 We conjecture the inversion relations for thermalized solvable interaction round the face (IRF)
 two dimensional lattice models. We base ourselves on an ansatz for the Baxterization described
 by the author in the 90's. We solve these inversion relations in the four main regimes of the models, to
 give the free energy of the models, in these regimes.
We use  the method used by Baxter in the calculation of the free energy of the hard hexagon model. We believe
 these results to be quite general, shared by most of the known IRF models. Our results apply equally
 well to solvable vertex models.
 Using the expression for the free energy we calculate the critical exponent $\alpha$, and from
 it the dimension of the perturbing (thermal) operator in the fixed point conformal field theory (CFT).
 We show that it matches either the coset ${\cal O}/{\cal G}$ or ${\cal G}/{\cal O}$, where $\cal O$ is the original CFT used 
 to define the model and $\cal G$ is some unknown CFT, depending on the regime. This agrees with 
 known examples of such models by Huse and Jimbo et al.
 
 \endpage

\mysec{Introduction.}

Two dimensional solvable lattice models offer a rich ground to study such phenomena as phase 
transitions, universality and mathematical applications in knot theory. For reviews see
\REF\Baxter{R.J. Baxter, ``Exactly solved models in statistical mechanics", Dover Publications
(1982).}
\REF\Wadati{M. Wadati, T. Deguchi and Y. Akutsu, Physics Reports 180 (4) (1989) 247.}
\r{\Baxter,\Wadati}.
These models also enjoy a strong connection with two dimensional conformal quantum filed theory
(CFT). See, e.g.,  the reviews
\REF\Ginsparg{P. Ginsparg, ``Applied conformal field theory", Les Houches lectures (1988).}
\REF\Francesco{P. Francesco, P. Mathieu and D. Senechal, ``Conformal field theory", Springer-Verlag,
Graduate texts in contemporary physics (1997).}
\r{\Ginsparg,\Francesco}.

Some time ago the author introduced a method to construct solvable interaction round the face
(IRF) from the data of an arbitrary CFT
\REF\Found{D. Gepner,  ``Foundation of rational quantum field theory", hep-th/9211100 (1992).}
\r\Found. We call such models IRF$({\cal O},h,v)$ where $\cal O$ is the defining CFT and
$h$ and $v$ are two primary fields in the theory.

A long standing question is what is the fixed point CFT of the so defined models and how it is related
to the original CFT $\cal O$. We solve this problem here by calculating the free energy of the
thermalized models. From this we deduce the critical exponent $\alpha$ and the dimension
of the perturbing field in the fixed point CFT.  

To compute the free energy we first need to thermalize the trigonometric ansatz of \r\Found.
This we do by calculating the two inversion relations for the general IRF models. Then
we thermalize the models by replacing the $\sin(u)$ function in the inversion relations
with the function $\theta_1(u,q)$, where $\theta_1$ is the standard elliptic function.
This agrees with all the models where the off--critical Boltzmann weights are known and
we conjecture that it is true in general. Thus we are in a position to solve exactly models 
for which the Boltzmann weights are not explicitly known.

We find that in the four main regimes of the IRF model the fixed point CFT is given
by a coset of the original theory. Namely, in regimes III and IV the fixed point CFT is
consistent with the coset model ${\cal G}/{\cal O}$ where $\cal G$ is some unknown CFT.
In regimes I and II, the fixed point CFT is ${\cal O}/{\cal G}$.

This  fixed point RCFT  is known exactly, in some cases. For example,
in the Andrews--Baxter--Forrester model
\REF\ABF{G.E. Andrews, R.J. Baxter and P.J. Forrester, J.Statist.Phys. (35) (1984) 193.}
\r\ABF, which is IRF$(SU(2)_k,[1],[1])$,
the fixed point field theory was determined to be the unitary minimal models,
which are the coset $SU(2)_{k-1}\times SU(2)_1/SU(2)_k$,  in regimes  III and IV
\REF\Huse{D.A. Huse, Phys.Rev.B 30 (1984) 3908.}
\r\Huse. In regimes I and II the critical CFT was identified as the parafermionic field theory
$SU(2)_k/U(1)$, which are the Fateev--Fateev  model
\REF\ZF{V.A. Fateev and A.B. Zamolodchikov, Sov.Phys.JETP 62 (1985) 215}
\r\ZF, by Jimbo et al.
\REF\Jimbo{M. Jimbo, T. Miwa and M. Okado, Nucl.Phys. B 275 (1986) 517.}
\r\Jimbo. 
Indeed, this agrees with our general result.
For the case of ${\cal O}=SU(N)_k$, and $h=v=$ fundamental, the fixed point field theory in regime III was shown to be
$SU(N)_{k-1}\times SU(N)_1/ SU(N)_k$ by Jimbo et al.
\REF\Jimbob{M. Jimbo, T. Miwa and M. Okado, Nucl.Phys. B 300 (1987) 74.}
\r\Jimbob. Again agreeing with our result for the fixed point CFT.

To compute the free energy, in all the four regimes, we follow the method used by Baxter \r\Baxter\ in the hard hexagon
model. Our results for the free energies agree with the hard hexagon case for ${\cal O}=SU(2)_3$,
in the four regimes.

\mysec{The inversion relations.}

We wish to study IRF lattice models based on the braiding matrix of conformal field theory (CFT). We fix
a conformal field theory $\cal O$ and  fixed primary fields in this theory $h$ and $v$. The IRF model is
denoted as IRF$({\cal O},h,v)$
following ref.
\r\Found, defined on a square lattice.
We assume that the boundary conditions are periodic.
Let $B_i$ be the braiding matrix in the RCFT which exchanges the
field $h$ with the field $v$
\REF\MS{G.W. Moore and N. Seiberg, Phys.Lett. B 212 (1988) 451.}\r\MS.
We define the operator
$$<a_1,a_2,\ldots, a_n| B_i|a_1^\prime,a_2^\prime,\ldots,a_n^\prime>=
B\left(\matrix{a_{i-1} & a_i\cr a_i^\prime & a_{i+1}\cr}\right) \prod_{m=1\atop m\neq i}^n\delta_{a_m,a_m^\prime},\e$$
where the matrix $B$ is the braiding matrix and it obeys the braiding relations, for $h=v$,
$$B_i B_{i+1} B_i=B_{i+1} B_i B_{i+1},\qquad B_i B_j=B_j B_i\ {\rm if\ } |i-j|>1.\e$$
The variables on the lattice $a_m$ and $a_m^\prime$ are some primary fields in the RCFT $\cal O$.

From the braiding matrix one can define the projectors,
$$P_i^b=\prod_{a=1\atop a\neq b}^n {B_i-\lambda_a\over \lambda_b-\lambda_a},\e$$
where $n$ is the number of eigenvalues of $B_i$ (called the number of blocks) and $\lambda_a$ are the eigenvalues. The projection operators obey the relations,
$$P_i^a P_i^b=\delta_{a,b} P_i^a,\qquad \sum_{a=1}^n P_i^a=1_i,\qquad \sum_{a=1}^n \lambda_a P_i^a=B_i,\e$$
where $1_i$ denotes the unit matrix. The eigenvalues of $B_i$ are \r\MS
$$\lambda_a=\epsilon_a  e^{\pi i(2\Delta_h-\Delta_a)},\e$$
where $\epsilon_a=\pm1$ according to whether the product is symmetric or anti--symmetric.

We define the fusion products of the field $h$ as,
$$h\cdot h=\sum_{a=0}^{n-1} \psi_a,\e$$
and
$$h\cdot \bar h=\sum_{a=0}^{n-1} \twidle\psi_a,\e$$
where $\bar h$ is the complex conjugate field of $h$, $n$ is the number of blocks and the order of the fields
is set in a certain way, which allows for the Yang--Baxter equation of the model. The order of the fields appears to be
that $\psi_{a+1}$ is contained in the fusion product of $\psi_a$ with the adjoint representation,
and similarly for $\twidle \psi_a$. 
(The fact that the number of blocks is the same in both eqs. (2.6, 2.7) is seen by computing the coefficient of the unit field in the fusion product $h\cdot h\cdot \bar h\cdot \bar h$ in two ways.)
In particular,
we set $\twidle\psi_0=[1]$ (the unit primary field) and $\twidle\psi_1=[{\rm adjoint}]$ (the adjoint representation, assuming
some quantum group structure). We denote the dimension of $\psi_a$ as $\Delta_a$ and similarly for
$\twidle\psi_a$ the dimension is $\twidle \Delta_a$.
We define the crossing parameters as,
$$\zeta_a={\pi\over2}(\Delta_{a+1}-\Delta_{a}),\e$$
and
$$\twidle \zeta_a={\pi\over2}(\twidle\Delta_{a+1}-\twidle\Delta_{a}),\e$$
where $a=0,1,\ldots, n-2$. We note that $\zeta_a,\twidle\zeta_a<\pi/2$, which will be important later.

In ref. \r\Found\  an ansatz for the trigonometric solution of the Yang Baxter
equation (YBE) was given. It is
$$R^{h,h}_i(u)=\sum_{a=0}^{n-1} f_a(u) P^{a}_i ,\e$$
where
$$f_a(u)=\left[ \prod_{j=1}^a \sin(\zeta_{j-1}-u) \right ] \left[ \prod_{j=a+1}^{n-1} \sin(\zeta_{j-1}+u)\right]\bigg/
\left[ \prod_{j=1}^{n-1} \sin(\zeta_{j-1})\right],\e$$
where $a=0,1,\ldots,n-1$.
Our ansatz is  that $R^{h,h}_i$ solves the Yang Baxter equation,
$$R_{i+1}(u) R_i(u+v) R_{i+1}(v)=R_i(v) R_{i+1}(u+v) R_i(u),\e$$
where we denoted by $R_i(u)$ instead of  $R^{h,h}_i(u)$, for shortness.  $R_i(u)$  is a trigonometric solution. Note that the YBE holds
only for a certain order of the fields in eqs. (2.6, 2.7). Similarly, we define $R^{h,\bar h}_i(u)$ by substituting
the crossing parameters $\twidle \zeta_a$ instead of $\zeta_a$ and substituting the
projection operators of the braiding matrix of $h$ with $\bar h$. $R^{h,\bar h}_i(u)$ also obeys some
sort of YBE,
$$R^{h,\bar h}_{i}(u) R_{i+1}^{h,\bar h}(u+v) R^{h,h}_{i}(v)=R^{h,h}_{i+1}(v) R^{h,\bar h}_{i}(u+v) R^{h,\bar h}_{i+1}(u),\e$$
The two YBE equations (2.12, 2.13) imply that the transfer matrices for $R^{h,h}_i(u)$ commutes with each other
for different spectral parameters $u$ and the same for $R^{h,\bar h}_i(u)$
\r\Baxter.

The $R$ matrices obey the first inversion relation which follows from eqs. (2.10, 2.11),
$$R^{h,h}_i (u) R^{h,h}_i(-u)=\rho(u) \rho(-u)\,1_i,\e$$
and 
$$R^{h,\bar h}_i (u) R^{h,\bar h}_i(-u)=\twidle \rho(u) \twidle \rho(-u)\, 1_i,\e$$
where 
$$\rho(u)=\prod_{r=0}^{n-2} \sin(\zeta_r-u)/\sin(\zeta_r),\e$$
and
$$\twidle \rho(u)=\prod_{r=0}^{n-2} \sin(\twidle \zeta_r-u)/\sin(\zeta_r),\e$$
where we changed by an irrelevant factor the normalization of $R^{h,\bar h}_i(u)$.

The second inversion relation is crossing. We shall denote again $R_i(u)$ by its matrix form. Then the crossing relation is, (as part of our conjectured ansatz), 
$$R^{h,\bar h}\pmatrix{d & c\cr a & b\cr}(u)=\left({\psi_a \psi_c\over \psi_b\psi_d}\right)^\half 
R^{h,h}\pmatrix{a & d \cr b & c \cr}(\lambda-u),\e$$
where $\lambda$ is the crossing parameter,
$$\lambda=\twidle \zeta_0=\pi \Delta_{\rm adjoint}/2,\e$$
where we used eq. (2.8). The crossing multipliers are
$$\psi_a=S_{a0}/S_{00},\e$$
where $S$ is the modular matrix
\r\Francesco.

We wish to thermalize now the IRF model. We do not know how to thermalize the Boltzmann weights.
So instead we will thermalize the inversion relations. We define the theta function,
$$\theta_1(u,q^2)=\sin u \prod_{n=1}^\infty (1-2 q^{2n} \cos 2u+q^{4n})(1-q^{2n}).\e$$
(This definition differs from the standard one by a factor of $2q^{1/4}$, which is irrelevant since we will
only encounter ratios of theta functions.) 

Now we conjecture that the thermalization of the first inversion relation, eq. (2.14), is given by replacing the
$\sin$ by the theta function $\theta_1$. We denote $\theta_1(u,q^2)$ by $\theta_1(u)$. Then, the thermalization of the $R$ matrix is
$$R^{h,h}_i (u) R^{h,h}_i(-u)=\rho(u) \rho(-u)\,1_i,\e$$
where
$$\rho(u)=\prod_{r=0}^{n-2} \theta_1(\zeta_r-u)/\theta_1(\zeta_r),\e$$
Similarly, we thermalize the second relation, eqs. (2.15, 2.18),
$$R^{h,\bar h}_i (u) R^{h,\bar h}_i(-u)=\twidle \rho(u) \twidle \rho(-u)\, 1_i,\e$$
where
$$\twidle \rho(u)=\prod_{r=0}^{n-2} \theta_1(\twidle \zeta_r-u)/\theta_1(\zeta_r).\e$$
Finally, the crossing relation eq. (2.18) remains the same for general $q$ except for the crossing multiplier,
eq. (2.20), whose explicit expression we will not need here. Note that for $q=0$ (the critical limit),
$\theta_1(u)=\sin u $ so we get the same inversion relations as before.

These conjectures can be indeed verified for many models for which we know the explicit Boltzmann
weights, e.g.,
\r{\Baxter,\Wadati}.

Next, we wish to define the free energy of the model. It is given by
$$\kappa=\lim_{N\rarrow \infty} Z^{1/N},\e$$
where $N$ is the number of lattice sites and $Z$ is the partition function calculated with $R^{h,h}$. The free energy is given as
usual by,
$$F=-k_B T \log \kappa,\e$$
where $k_B$ is Boltzmann constant and $T$ is the temperature.

Now, since the transfer matrices commute for different spectral parameters $u$, the inversion
relations translate to equations of $\kappa(u)$ (fixing some $q$),
$$\kappa(u) \kappa(-u)=\rho(u) \rho(-u),\e$$
and
$$\kappa(\lambda+u) \kappa(\lambda-u)=\twidle \rho(u)\twidle\rho(-u).\e$$
In deriving the last equation, we used the fact that the crossing multipliers cancel when calculating the
partition function.

Actually, the inversion relations, eqs. (2.28, 2.27), remain the same under the substitution $\zeta_i\rarrow -\zeta_i$,
or $\twidle\zeta_i\rarrow -\twidle\zeta_i$. Thus, we take instead of the crossing relations their
absolute values, $\zeta_i\rarrow |\zeta_i|$ and $\twidle\zeta_i\rarrow |\twidle\zeta_i|$.

We also find it convenient to change the second inversion relation by substituting $u\rarrow \lambda-u$.
The second inversion relation then becomes,
$$\kappa(u) \kappa(2\lambda-u)=\prod_{r=0}^{n-2} \left[ \theta_1(\twidle \zeta_r-\lambda+u)
\theta_1(\twidle\zeta_r+\lambda-u)\right].\e$$ 

\mysec{Regimes III.}

Our aim now is to solve the inversion relations eqs. (2.28, 2.30) and to calculate the free energy. 
We assume first that the model is in regime III. This is defined by
$$0<q^2<1,\e$$
$$0<u<d,\e$$
where $d=\min_i \zeta_i,\twidle \zeta_i$,  and
$q^2=\exp(-\epsilon)$.
It is convenient to use the modular transformation of the theta function. This is defined by \r\Baxter
$$\theta_1(u,e^{-\epsilon})=\half \left({2\pi\over\epsilon}\right)^\half
\exp\left[ {\epsilon\over8}-{\pi^2\over2\epsilon}+{2u(\pi-u)\over\epsilon}\right]
f(e^{-4\pi u/\epsilon},e^{-4\pi^2/\epsilon}),\e$$
where $f$ is defined by,
$$f(w,q)=\prod_{n=1}^\infty (1-q^{n-1} w) (1-q^n w^{-1})(1-q^n).\e$$
We find it convenient to redefine,
$$\twidle\kappa(u)=e^{2\delta u/\epsilon} e^{2(n-1)  u^2/\epsilon}\kappa(u),\e$$
where
%$$\delta=\sum_{r=0}^{n-2} [\lambda^2-\twidle\zeta_r(\pi-\twidle\zeta_r) +\zeta_r(\pi-\zeta_r)]/\lambda.\e$$ %
$$\delta={1\over\lambda} \sum_{r=0}^{n-2} [\zeta_r(\pi-\zeta_r) -\pi\twidle\zeta_r].\e$$
Then,  the first inversion relation, eq. (2.28, 2.23), becomes,
$$\twidle k(w) \twidle k(w^{-1})=\eta(w),\e$$
where
$$\eta(w)=\prod_{r=0}^{n-2} {f(z_r w) f(z_r w^{-1})\over f(z_r)^2},\e$$
where 
$$w=\exp(-4\pi u/\epsilon),\e$$
$$z_i=\exp(-4\pi \zeta_i/\epsilon),\e$$
$$\twidle z_i=\exp(-4\pi \twidle \zeta_i/\epsilon),\e$$
$$\twidle q=e^{-4\pi^2/\epsilon},\e$$
and we denoted for brevity $f(w)$ for $f(w,\twidle q)$.
The second inversion relation, eq. (2.30), becomes,
$$\twidle\kappa( w) \twidle\kappa(x^2  w^{-1})=\twidle \eta(w),\e$$
where 
$$x=\exp(-4\pi \lambda/\epsilon)=\twidle z_0,\e$$
and
$$\twidle\eta(w)=\prod_{r=0}^{n-2} {f(\twidle z_r w/x) f(\twidle z_r x  w^{-1})\over f(z_r)^2}.\e$$

We wish to solve now for the free energy, using the two inversion relations. 
For this purpose we assume that 
$\log(\twidle k(w))$ is analytic in the annulus containing the point $w=1$ and the point $w=x$. We analytically continue
$k(u)$ to $-\lambda <u<0$ and to $0<u<\lambda$. This assumption is justified by considering 
explicit models, e.g., the hard hexagon model \r\Baxter. 

So we expand,
$$\log[\twidle k(w)]=\sum_{m=-\infty}^\infty c_m w^m,\e$$
where the summation is convergent in the annulus containing $1$ and $x$ . In the neighborhood of
$1$, eqs. (3.7, 3.8), gives,
$$\log\eta=d_0+\sum_{m=1}^\infty d_m(w^m+w^{-m}),\e$$
where $m>0$,
$$d_0=\sum_{r=0}^{n-2} -2\log [f(z_r)/\phi(\twidle q)],\e$$
where
$$\phi(\twidle q)=\prod_{m=1}^\infty (1-\twidle q^m),\e$$
and
$$d_m=-\sum_{r=0}^{n-2} (z_r^m+\twidle q^m z_r^{-m})/[m(1-\twidle q ^m)].\e$$
We assume $d_{-m}=d_m$.
Similarly, the second inversion relation, eqs. (3.13--3.15), becomes, for $m\neq 0$,
$$\log\twidle \eta= \sum_{m=-\infty}^\infty d_m^\prime  w^m,\e$$
where 
$$d_0^\prime=d_0,\e$$
$$d_m^\prime=-\sum_{r=0}^{n-2} (\twidle z_r^m/x^m+\twidle q^m \twidle z_r^{-m}x^{-m})/[m(1-\twidle q ^m)].\e$$

Taking the logarithms of eqs. (3.13, 3.7) and equating coefficients, we find,
$$c_m+c_{-m}=d_m,\e$$
$$ c_m+x^{-2 m} c_{-m}=d_m^\prime,\e$$
for $m\geq 0$.
The solution of these two equations is,
$$c_0=\half d_0,\e$$
$$c_m=(-x^{2m} d_m^\prime+d_m)/(1-x^{2 m}),\e$$
for $m\neq 0$.
This completes the calculation of the free energy in regime III. Indeed 
$$\log \twidle \kappa(w)=\sum_{m=-\infty}^\infty  c_m w^m,$$ 
converges in an annulus containing the points $1$ and $x$, as we assumed.
Note that $|\twidle\zeta_r|\leq \lambda$ always, for all $r$, which is needed to show convergence, which we checked 
in many models, but we do not have a general proof for this fact.

We wish to calculate now the critical exponent $\alpha$. This is defined as the singularity of the
free energy,
$$\kappa_{\rm singular} \propto |T-T_c|^{2-\alpha},\e$$
where $T_c$ is the critical temperature. We have that $\twidle q=\exp(-4\pi^2/\epsilon)$.
Since $c_m$ is divided by $1-x^{2m}$ this means that it becomes a theta function at the
modulus $x^2$, when summed back. Since $\lambda=\pi\twidle \Delta_0/2$, eq. (2.9), we may write this 
modulus as
$$x^2=\twidle q^{\twidle \Delta_0}.\e$$
Under a modular transformation, $x^2$ becomes,
$$(q^2)^{1/\twidle\Delta_1}.\e$$
Since $q^2\propto |T-T_c|$ this implies that
$$ \kappa_{\rm singular}\propto |T-T_c|^{1/\twidle\Delta_0},\e$$
and so,
$$2-\alpha={1\over \twidle\Delta_0}.\e$$
The dimension of the perturbing operator at the critical conformal field theory is given by
\r\Ginsparg,
$$\Delta_p={1-\alpha\over 2-\alpha}=1-\twidle\Delta_0.\e$$
Now, since  we assume that $\twidle\psi_1$ is the adjoint operator, in quantum group models,
as can be see by considering various models, the dimension of the perturbing
field is seen to be,
$$\Delta_p=1-\Delta_{\rm adjoint},\e$$
where $\twidle\Delta_0=\Delta_{\rm adjoint}$.
\foot{An exception to this is the hard hexagon model, where $\kappa(w)$ is not singular in regime III,
as a result of
cancelations \r\Baxter.} 
Such a field appears in the coset theory,
$${\cal C}={{\cal G}\over {\cal O}},\e$$
where $\cal O$ is the original CFT used to define the model and $\cal G$ is some CFT model,
where we take the currents in $\cal G$ and the adjoint representation in $\cal O$. Thus,
we conjecture that the fixed point of Regime III is given by the coset $\cal C$. In many cases where
the fixed point theory was calculated explicitly,
 this was indeed shown to be the case. For example, the RCFT $\cal O$  of the ABF
 model \r\ABF\ is $SU(2)_k$ WZW model. The fixed point in Regime III is the
 $k+1$ minimal model 
\r\Huse,
 defined by the coset
 $${\cal C}={SU(2)_{k-1}\times SU(2)_1\over SU(2)_k},\e$$
 consistent with eq. (3.35).
 
 \mysec{Regime II.}
Let us consider regime II. This is defined by
$$0<q<1,\qquad 0>u>-d.\e$$
The first inversion relation remains the same, eqs. (2.28, 2.23). The second inversion relation becomes,
$$\kappa(u)\kappa(2\lambda-\pi-u)=\prod_{r=0}^{n-2} \theta_1(\twidle\zeta_r-\lambda+u)
\theta_1(\twidle\zeta_r+\lambda-\pi-u)/\theta_1(\zeta_r)^2,\e$$
where the theta function is invariant under the shift by $\pi$.
We define,
$$\twidle\kappa(u)=e^{\delta u/\epsilon} e^{2(n-1) u^2/\epsilon}\kappa(u),\e$$
where 
$$\delta=\sum_{r=0}^{n-2} \left[ 
2\pi^2-4\pi\twidle\zeta_r+4\zeta_r(\pi-\zeta_r)\right]\bigg/(2\lambda-\pi).\e$$
Then the second inversion relation, eq. (4.2), becomes,
$$\twidle\kappa(w) \twidle\kappa(w_0^2/w)=\twidle\eta(w) =\prod_{r=0}^{n-2}  \left[ f(\twidle z_r x^{-1} w)
f(\twidle z_r x \twidle q^{-1} w^{-1})/f(z_r)^2 \right],\e$$
where
$$w_0^2=\twidle q^{-1} x^2,\e$$
where $\twidle q$, $x$, $z_r$ and $\twidle z_r$  are given by eqs. (3.12, 3.14, 3.10, 3.11). 
We assume now that 
$$\log[w^{\mu}\twidle\kappa(w)]\e$$
is analytic in an annulus $a<|w|<b$ containing the points $w=1$ and $w=w_0$. This is an analytic
continuation of $w$ to $w<1$ and $w>w_0$ even though the regime is defined for a subset of these 
values.

Then we can calculate 
$$\log\twidle\eta(w)=\sum_{m=-\infty}^\infty d_m^\prime w^m,\e$$
and
$$d_m^\prime=\sum_{r=0}^{n-2} [\twidle z_r^m x^{-m}+\twidle q^{-m}(\twidle z_r x \twidle q^{-1})^{-m}]/
[m(1-\twidle q^{-m})],\e$$
$$d_0^\prime=d_0+\sum_{r=0}^{n-2} \log(\twidle z_r^2 \twidle q^{-1}).\e$$
Thus,
$$\mu=(d_0^\prime-d_0)/(2\log w_0)={\twidle\Delta_{n-1}\over\Delta_{\rm adjoint}},\e$$
and 
$$\log[w^\mu \twidle\kappa(w)]=\sum_{m=-\infty}^\infty c_m w^m,\e$$
where
$$c_m=(d_m-w_0^{2m} d_m^\prime)/(1-w_0^{2m}),\e$$
for $m\neq0$, and
$$c_0=\half d_0.\e$$
Indeed the series, eq. (4.12), for $\twidle \kappa$ converges in the annulus containing $w=1$ and $w=w_0$.

The free energy $\log[w^\mu \twidle \kappa(w)]$ is given by the series
$$c_m=a_m/(1-w_0^{-2}),\e$$
where $a_m$ is some function. Thus it is given by a theta function with the modulus
$w_0^{-2}$ which is
$$p=w_0^{-2}=e^{-4\pi^2(1-\twidle\Delta_0)/\epsilon}=\twidle q^{1-\twidle \Delta_0},\e$$
under a modular transformation, as in eq. (3.3), we find
$$\kappa_{\rm singular}=(q^2)^{1/(1-\twidle\Delta_0)},\e$$
where $q^2$ is proportional to $|T-T_c|$. Thus we find for the exponent $\alpha$,
$$2-\alpha=1/(1-\twidle\Delta_0).\e$$
For a quantum group model $\twidle\Delta_0$ is the dimension of the adjoint representation,
conjecturally. Thus the dimension of the perturbing thermal operator is
$$\Delta_p={1-\alpha\over 2-\alpha}=\twidle\Delta_0=\Delta_{\rm adjoint}.\e$$
An operator with such a dimension appears in the Coset model,
$${\cal C}={{\cal O}\over {\cal  G}},\e$$
where $\cal O$ is the RCFT used to define the model and $\cal G$ is some CFT model. 
Thus we conjecture that $\cal C$ is the fixed point theory in regime II.
The
perturbing operator is the adjoint representation in $\cal O$ and the unit in $\cal G$.
For example, for the ABF model in regime II, where ${\cal O}$ is $SU(2)_k$ WZW model,
the fixed point theory is
the parafermionic theory \r\Jimbo,
$${\cal C}={SU(2)_k\over U(1)}.\e$$

\mysec{Regime IV.}

Let us get now to regimes IV and I. Regime IV is defined by
$$-1<q^2<0,\qquad 0<u<d,\e$$
where $d=\min_i \zeta_i,\twidle \zeta_i$. We use the inverse modulus \r\Baxter,
$$\theta_1(u,-e^{-\epsilon})=-\half \left({\pi\over\epsilon}\right)^\half 
\exp\left[ {\epsilon\over 8}-{\pi^2\over 8\epsilon}-{u(2u+\pi)\over\epsilon}\right]
f(e^{2\pi u/\epsilon},-e^{-\pi^2/\epsilon}),\e$$
where, as before,
$$f(w,q)=\prod_{n=1}^\infty (1-q^{n-1} w)(1-q^n w^{-1}) (1-q^n).\e$$
We define
$$q^2=-\exp(-\epsilon),\quad \twidle q=-e^{-\pi^2/\epsilon},\e$$
$$x=-\exp(-\pi \lambda/\epsilon),\quad z_r=-\exp(-\pi\zeta_r/\epsilon),\quad \twidle z_r=
-\exp(-\pi\twidle\zeta_r/\epsilon),\e$$
$$w=\exp(2\pi u/\epsilon),\e$$
where the crossing parameters were defined in eqs. (2.8, 2.9).

In regime IV we write the two inversion relations as
$$\kappa(u)\kappa(-u)=\prod_{r=0}^{n-2} \left[ \theta_1(-\zeta_r-u)\theta_1(-\zeta_r+u)/\theta_1(-\zeta_r)^2
\right],\e$$
$$\kappa(u)\kappa(2\lambda-u)=\prod_{r=0}^{n-2}\left [\theta_1(\twidle \zeta_r-\lambda+u)
\theta_1(\twidle\zeta_r+\lambda-u)/\theta_1(-\zeta_r)^2\right],\e$$
which are eqs. (2.28, 2.30).
We define
$$\twidle\kappa(u)=e^{\delta u/(2\lambda \epsilon)} e^{2(n-1)u^2/\epsilon} \kappa(u),\e$$
where
$$\delta=\sum_{r=0}^{n-2} 2 \pi \twidle \zeta_r+2\zeta_r(\pi-\zeta_r).\e$$
then, the first inversion relation (for regimes I and IV) becomes,
$$\twidle\kappa(w)\twidle\kappa(w^{-1})=\eta(w)=\prod_{r=0}^{n-2} {f(z_r^2 w) f(z_r^2/w)\over
f(z_r^2)^2}.\e$$
The second inversion relation, for regime IV, becomes,
$$\twidle\kappa(w) \twidle\kappa(x^{-4}/w)=\twidle \eta(w)=
\prod_{r=0}^{n-2} \left[ f(w x^2\twidle z_r^{-2}) f(w^{-1} \twidle z_r^{-2} x^{-2} )/f(z_r^2)^2\right],\e$$
where we define the crossing parameter,
$$w_0=x^{-2}.\e$$
We denoted for brevity $f(w,\twidle q)$ as $f(w)$.

As before, we assume that $\log(w^\mu \twidle\kappa(w))$ is analytic in an annulus
$a<|w|<b$ containing $w=1$ and $w=w_0$. Thus we expand
$$\log(\eta(w))=\sum_{m=-\infty}^\infty d_m w^m,\e$$
and we find (in regimes I and IV)
$$d_m=-\sum_{r=0}^{n-2} { z_r^{2 m}+z_r^{-2m} \twidle q^m \over m(1-\twidle q^m)},\e$$
$$d_0=-2\sum_{r=0}^{n-2} \log [f(z_r^2)/\phi(\twidle q)],\e$$
where
$$\phi(\twidle q)=\prod_{m=1}^\infty (1-\twidle q)^m.\e$$

Similarly we expand
$$\log(\twidle \eta(w))=\sum_{m=-\infty}^\infty d_m^\prime w^m,\e$$
where
$$d_m^\prime=\sum_{r=0}^{n-2} {(x^2 \twidle z_r^{-2})^m+(\twidle z_r^{-2} x^{-2})^{-m} \twidle q^{-m}
\over m (1-\twidle q^{-m})},\e$$
for $m\neq0$, and
$$d_0^\prime=d_0+2\mu\log w_0,\e$$
where
$$\mu=\sum_{r=0}^{n-2} \log(\twidle z_r^{-4})/\log (x^{-4})={\twidle\Delta_{n-1}\over \Delta_{\rm adjoint}}.\e$$

Thus, as before, eq. (3.27), we find 
$$\log(w^\mu  \twidle \kappa(w))= \sum_{m=-\infty}^\infty c_m w^m,\e$$
where
$$c_m={(d_m-w_0^{2m} d_m^\prime )\over (1-w_0^{2m})},\e$$
for $m\neq 0$ and
$$c_0=\half d_0.\e$$ 
We note that indeed this series converges in the annulus containing $w=1$ and $w=w_0$.

Let us compute now the critical exponent $\alpha$. The series for $c_m$, eq. (5.23), is an expression
for a theta function with the modulus,
$$w_0^{-2}=x^4=e^{-2\pi^2 \Delta_{\rm adjoint} /\epsilon}=\twidle q^{2\Delta_{\rm adjoint}},\e$$
(we ignore the overall sign).  
The expression for $\twidle \kappa$ includes factors of $\theta_4(u,-\twidle q^2)$.
The function $\theta_4$ is defined by
$$\theta_4(u,q^2)=\prod_{n=1}^\infty (1-2 q^{2n-1} \cos 2u+q^{4n-2})(1-q^n),$$
and it satisfies the conjugate modulus relation
$$\theta_4(u,e^{-\epsilon})=\left({2\pi\over\epsilon}\right)^\half  \exp\left[{-\pi^2\over 2\epsilon}+{2u(\pi-u)\over\epsilon}\right] f(-e^{-4\pi u/\epsilon},e^{-4\pi^2/\epsilon}).$$
Denoting $p=-q^2$, we find that the inverse modulus transformation gives the modulus,
$$p^{2/\Delta_{\rm adjoint}}.\e$$
(Note that we get a factor of $4$ in the modulus as a result of the inverse modulus relation.)
Now, since the function $\theta_4$ includes powers of $q=p^\half$ and defining $p=t$ where $t$ is the temperature,
we find that  
$$\kappa_{\rm singular}\propto t^{1/\Delta_{\rm adjoint}},\e$$
and
$$2-\alpha=1/\Delta_{\rm adjoint}.\e$$
Thus the dimension of the perturbing (thermal) operator in the fixed point RCFT is
$$\Delta_p={1-\alpha\over 2-\alpha}=1-\Delta_{\rm adjoint}.\e$$
This is the same dimension as in Regime III and we conclude that it is the other side 
of the same fixed point. The RCFT in Regime IV is thus conjectured as the coset model
$${\cal C}={{\cal G}\over \cal O},\e$$
where $\cal O$ is the original CFT used to define the model and $\cal G$ is some unknown CFT.

\mysec{Regime I.}

Let us turn now to regime I. It is defined by
$$-1<q<0,\qquad 0>u>-d,\e$$
The first inversion relation is the same as in regime IV, eq. (5.7). For the second inversion relation
we take as in regime II,
$$\kappa(u)\kappa(2\lambda-\pi-u)=\prod_{r=0}^{n-2} \theta_1(\twidle\zeta_r-\lambda+u)
\theta_1(\twidle\zeta_r+\lambda-\pi-u).\e$$
We define as in regime IV, $q$, $\twidle q$, $x$, $\twidle z_r$ and $w$, eqs. (5.4--5.6). We further define
$$\twidle\kappa(u)=e^{2(n-1) u^2/\epsilon} e^{\delta u/[\epsilon(2\lambda-\pi)]} \kappa(u),\e$$
where
$$\delta=\sum_{r=0}^{n-2} 2\zeta_r(\pi-2\zeta_r)-\pi^2+2\pi \twidle\zeta_r.\e$$
Then the second inversion relation, eq. (6.2), then becomes
$$\twidle\kappa(w) \twidle\kappa(w_0^2/w)=\prod_{r=0}^{n-2} f(w x^2 \twidle z_r^{-2}) 
f(w^{-1} \twidle z_r^{-2} x^{-2} \twidle q^2)=\twidle \eta(w),\e$$
where
$$w_0=-x^{-2} \twidle q,\e$$
and $f(w,\twidle q)$ was defined in eq. (5.3).

Now, we need to expand as before
$$\log\twidle \eta(w)=\sum_{m=-\infty}^\infty d_m^\prime w^m,\e$$
and we find,
$$d_0^\prime=d_0,\e$$
where $d_0$ is given by eq. (5.16). Also,
$$d_m^\prime=-\sum_{r=0}^{n-2} {(x^2\twidle z_r^{-2})^m+(\twidle z_r^{-2} x^{-2} \twidle q^2)^{-m}
\twidle q^m \over m (1-\twidle q^m)},\e$$
for $m\neq0$.
So, as before,
$$\log\twidle\kappa(w)=\sum_{m=-\infty}^\infty c_m w^m,\e$$
where
$$c_0=\half d_0,\e$$
and
$$c_m=-{w_0^{2m} d_m^\prime-d_m\over 1-w_0^{2m} },\e$$
as before, for $m\neq0$.
The series, eq. (6.10), indeed converges in the annulus containing $w=1$ and $w=w_0$.

Now, we wish to compute the exponent $\alpha$. The expression for $c_m$ is a theta function
with the modulus 
$$w_0^2=\twidle q^2 x^{-4},\e$$
which can be written as
$$w_0^2=\twidle q^{2(1-\Delta_{\rm adjoint})},\e$$
where we used the definition of $x$ and $\lambda$, eqs. (5.5, 2.9).
Thus, the modulus of the theta function has the modulus
$$(\twidle q^2)^{2(1-\Delta_{\rm adjoint})}.\e$$
Now, since the expression for the theta function contains $\theta_4$,  at positive modulus,
the same discussion as in section (5) shows that
$$2-\alpha={1\over 1-\Delta_{\rm adjoint}},\e$$
and the dimension of the perturbing field is
$$\Delta_p={1-\alpha\over 2-\alpha}=\Delta_{\rm adjoint}.\e$$
This is the same dimension as in regime II and we conclude that it is the other side of  the same
phase transition with the critical theory given as in regime II by the coset
$${\cal C}={{\cal O}\over {\cal G}},\e$$
where $\cal O$ is the original theory used to define the model and $\cal G$ is some unknown
CFT. 

This concludes the expression for the free energy in all the four regimes. We can check our results 
for the hard hexagon model which is IRF$(SU(2)_3,[1])$. There $\lambda=\pi/5$,
$n=2$, $\zeta_0=\twidle \zeta_0=\lambda$. We find a complete agreement with the results
of Baxter for the free energy \r\Baxter, p. 425.

\mysec{Discussion.}

Some other two dimensional solvable lattice models are the vertex models. These are defined
by some CFT $\cal O$ and some representation in them $h$. The Boltzmann weights are elements
of End$(V\times V)$ where $V$ are the weights of the representation $h$. As was discussed in refs.
\REF\Bk{D. Gepner, Nucl.Phys. B 958 (2020) 115116.}
\REF\Dk{V. Belavin, D. Gepner and H. Wenzl, Nucl.Phys. B 959 (2020) 115160.}
\r{\Bk,\Dk}, the Baxterization of the models is exactly the same formula as the IRF models, eq. (2.10, 2.11).
For the case of $SU(2)$ this was described in ref.
\REF\Pasquier{V. Pasquier, Commun.Math.Phys. 118 (1988) 355.} \r\Pasquier.
Thus, our calculation of the free energy holds equally well for these models, with the parameters
$\zeta_i$ and $\twidle \zeta_i$ given by eqs. (2.8, 2.9). We assume here periodic boundary conditions.
The free energy then obeys the same inversion relations, eqs. (2.28, 2.30). Thus also the solution 
for the free energy is 
the same as described in this paper.

Another interesting point is that in the theories $\cal C$ define integrable models when perturbed by
the thermal operator. This may be interesting for the building of new integrable models of 
massive quantum field theories.

\refout

\bye

\bye

%% file: mydef.tex
\def\e{\adveq\eqno{\rm (\chapterlabel.\the\equanumber)}}
\def\mysec#1{\equanumber=0\chapter{#1}}
\def\adveq{\global\advance\equanumber by 1}
\def\myeq{{\rm \chapterlabel.\the\equanumber}}
\def\rarrow{\rightarrow}

\def\semidirect{\mathrel{\raise0.04cm\hbox{${\scriptscriptstyle |\!}$
\hskip-0.175cm}\times}}

%\define\semidirect{\propto}

\def\ref#1{$^{[#1]}$}

\def\r#1{$[\rm#1]$} 
\def\twidle{\tilde}

\def\e{\adveq\eqno{\rm (\chapterlabel.\the\equanumber)}}
\def\mysec#1{\equanumber=0\chapter{#1}}
\def\adveq{\global\advance\equanumber by 1}
\def\myeq{{\rm \chapterlabel.\the\equanumber}}
\def\rarrow{\rightarrow}

\def\semidirect{\mathrel{\raise0.04cm\hbox{${\scriptscriptstyle |\!}$
\hskip-0.175cm}\times}}

%\define\semidirect{\propto}

\def\ref#1{$^{[#1]}$}

\def\r#1{$[\rm#1]$} 
\def\twidle{\tilde}

\def\half{{1\over2}}